\documentstyle[preprint,aps]{revtex}

\begin{document}


\title{Monte Carlo Study of Correlations Near the Ground State of the
       Triangular Antiferromagnetic Ising Model}

\author{Jesper Lykke Jacobsen$^1$ and Hans C.~Fogedby$^{1,2}$}

\address{$^1$Institute of Physics and Astronomy, University of Aarhus,
         DK-8000 Aarhus C, Denmark \\
         $^2$Nordita, Blegdamsvej 17, DK-2100, Copenhagen {\O}, Denmark}

\date{\today}

\maketitle


\begin{abstract}
  We study the spin-spin correlation function in or near the $T=0$
  ground state of the antiferromagnetic Ising model on a triangular
  lattice. At zero temperature its modulation
  on the sublattices gives rise to two Bragg peaks in the structure factor,
  and a known expression for the algebraic decay of correlations enables us to
  examine the form of the diffusive scattering. We do so by means of a
  comparison between exact results and data calculated using standard Monte
  Carlo techniques. At non-zero temperatures the finite correlation length
  alters this form, and we account for the change by proposing a generalisation
  of the zero temperature pair correlation function. The size dependence of
  our simulation data is investigated through a novel finite-size
  scaling analysis where $t = \mbox{\rm e}^{-2/T}$ is used as the
  temperature parameter.

  \vspace{0.5cm}
  {\bf PACS-numbers:} 64.60.Fr, 64.70.Rh, 75.10.Hk, 75.40.Mg

  \vspace{0.5cm}
\end{abstract}




Since Onsager's evaluation of the free energy for the Ising-model on a
rectangular lattice \cite{Onsager} the corresponding model on a triangular
lattice has served as an afterthought requiring some modification of the
analysis \cite{Wannier}.
The Triangular Antiferromagnetic Ising (TAI) model has also attracted
attention in its own right, it being a simple manifestation of a fully
frustrated system \cite{Kobe}. Very recently, the interest in the TAI
model has been revived through its near-perfect experimental
realisation in the yavapaiite layered structure of anhydrous alums
such as RbFe(SO$_4$)$_2$ \cite{Bramwell}.

Mathematically the isotropic TAI model to be studied in this paper is
defined by the Hamiltonian
\begin{equation}
  {\cal H} = J \sum_{\langle i,j \rangle} s_i s_j,
  \label{Tai}
\end{equation}
where $J>0$ is the antiferromagnetic exchange coupling and
$s_i = \pm 1$ are Ising spins on the triangular lattice depicted in
Fig.~\ref{Lattice}. The label $\langle i,j \rangle$ indicates a sum over
nearest neighbour pairs each pair being counted only once.
Frustration in the ground state arises from the
inability of the system to simultaneously satisfy the ``local packing rule''
that three spins on an elementary triangle are pairwise each other's nearest
neighbour, and the ``global packing constraint'' that, in order to minimise
energy, the ground state must have as many antiferromagnetically satisfied
nearest neighbour bonds as possible. In other words, it is impossible to
orient three spins in a pairwise antiparallel fashion.
It can thus be shown \cite{Wannier} that the ground state is
macroscopically degenerate with a finite entropy per spin
$s_0 = \frac{2}{\pi} k_B \int_0^{\pi/3} \ln (2 \cos \omega) \,\mbox{\rm d} \omega
     \simeq 0.323 \, k_B.$

In Wannier's original approach \cite{Wannier} the method of Onsager
(diagonalisation of the transfer matrix using representation theory on an
associated Lie algebra) was modified to include the diagonal interactions
of the TAI model. Due to the technical complexity of this method much work was
done to achieve a simplification of the algebra involved (see references in
\cite{Schultz}), the result being the now well-known technique of reducing
the problem to chains of interacting fermions \cite{Schultz}.

At the same time an alternative scheme, known as the combinatorial method,
began to emerge. Using topological theorems Kasteleyn \cite{Kast61,Kast63}
evaluated the configurational generating function of the problem covering
a lattice with dimers. The success of this approach relies on the fact that the
combinatorics can be lumped into a mathematical entity known as the Pfaffian,
which is simply the square root of a determinant. The Ising problem is
then addressed by counting dimer configurations on a modified lattice, in
which every vertex has been substituted by a suitably oriented polygon in
order to ensure a correct counting of the states summed over in the Ising
partition function \cite{Kast63,Montroll}.

This scheme (as opposed to the algebraic method) is easily augmented to
include the evaluation of correlation functions \cite{Montroll} expressed
as Toeplitz determinants. Using asymptotic properties for these objects,
Stephenson \cite{Steph64} provided the long-distance behaviour of the two-spin
correlations along the three main directions for the TAI model
with \cite{Steph70b} or without \cite{Steph70a} anisotropy.

A recent interest
\cite{Blote82,Blote84,Blote91,Blote93,NohKim94,NohKim95,Queiroz}
has been taken in generalising the Hamiltonian (\ref{Tai})
to include ferromagnetic next-nearest neighbour interactions as well as
anisotropy. Such additional couplings tend to lift the ground state
degeneracy, since, dividing the triangular lattice into the usual three
sublattices, next-nearest neighbour pairs belong to the same of these
sublattices (see Fig.~\ref{Lattice}).
By applying mappings which are constructed to automatically
satisfy the antiferromagnetic nearest neighbour constraint, the ground state
ensemble can be investigated within the context of solid-on-solid (SOS)
\cite{Blote82,Blote84,Blote93} or domain-wall \cite{NohKim94,NohKim95} models.

In this paper we shall concern ourselves with the nature of the correlations in
the low-temperature TAI model. Despite the fact that the abundance of
entropy prevents the emergence of a long-range order, even at zero temperature
\cite{Wannier}, the existence of some kind of short-range order or pattern
formation should be evident from a visual inspection of Fig.~\ref{Pattern},
and the $T=0$ state is indeed a critical one as witnessed by the algebraic
decay of correlations \cite{Steph70a}.

We shall study this ordering by means of the structure factor
\begin{equation}
  S({\bf p}) = \sum_{\bf r} e^{i {\bf p} \cdot {\bf r}}
               \langle s_{\bf 0} s_{\bf r} \rangle,
\end{equation}
which is proportional to the cross section for quasi-elastic single scattering
\cite{Nielsen}. When scattering, for example, neutrons from a rare-gas
monolayer adsorbed on graphite \cite{Schick81} the ordering is directly
displayed through the positions and shapes of peaks in the structure factor.

By Fourier transforming the known $T=0$ expression for the pair-correlation
function \cite{Steph70a} we find good agreement with numerical data
for the structure factor as calculated from standard Monte Carlo (MC) data.
Also, simulations performed at non-zero temperature render unchanged values for
the Bragg points but a broadening of the line shape, which makes us suggest
a generalisation of the expression for the pair-correlation function to the
$T>0$ case. Finally, the r\^{o}le of finite-size effects in the MC data is
discussed by means of a finite-size scaling analysis where $t =
\mbox{\rm e}^{-2/T}$ is used as the temperature parameter.


\vspace{0.5cm}

Stephenson \cite{Steph70a} made an asymptotic expansion of the spin-spin
correlation function for the $T=0$ TAI model
\begin{equation}
  \langle s_{\bf 0} s_{\bf r} \rangle \sim
  \frac{\cos \left( \frac{2\pi r}{3} \right)} {\sqrt{r}}
  \: \mbox{\rm for} \: r \gg 1,
  \label{Steph}
\end{equation}
valid along the three main directions of the lattice.
Since the derivation of this result depends heavily on certain
asymptotic properties of the Toeplitz determinants involved, a
similarly stringent result valid for arbitrary lattice directions
could hardly be worked out along these lines. However the symmetry
between the tree sublattices suggests that the general expression is
formed by replacing the factor $\cos(\frac{2\pi r}{3})$
with a weight factor having the value $+1$ when $s_{\bf 0}$ and $s_{\bf r}$
are on the same sublattice, and $-\frac{1}{2}$ otherwise \cite{Blote84}.

Recently it has been demonstrated that a staggered field on the
sublattices can be mapped onto a period-6 spin wave operator within
the context of an equivalent SOS-model \cite{Blote84}. By
renormalisation the latter is reduced to a Gaussian model thus
allowing for a determination of the associated critical exponent as
$X_6^{\rm{(s)}} = \frac{1}{4}$ which in in nice agreement with the algebraic
decay $\sim r^{-2 X_6^{\rm{(s)}}}$ in Eq.~(\ref{Steph}). Although this
of course corroborates the original assumption of a staggered field we
are not aware of anyone having formerly verified it through a direct
study of the TAI model.

Defining ${\bf q}_{1,2} = (\pm \frac{4\pi}{3},0)$ it is easily checked that
$\frac{1}{2} \left(
  e^{-i {\bf q}_1 \cdot {\bf r}} + e^{-i {\bf q}_2 \cdot {\bf r}} \right)$
is the wanted weight factor, and the structure factor is evaluated as
\begin{equation}
  S({\bf p}) \sim \sum_{j=1}^2 \int \mbox{\rm d}^2r
    e^{i({\bf p} - {\bf q}_j) \cdot {\bf r}} f(r),
  \label{TwinPeaks}
\end{equation}
where we have introduced the radially symmetric function
$f(r) = 1/\sqrt{r}$ for
the purpose of later generalisation. It is seen that the r\^{o}le of the
weight factor is to produce two Bragg peaks at wave vectors ${\bf q}_1$ and
${\bf q}_2$, as shown in Fig.~\ref{Brillouin}. Interestingly, these
positions are also predicted by a simple mean field theory \cite{Schick81},
pleading no knowledge of such exact results as Eq.~(\ref{Steph}). Furthermore,
the knowledge of an algebraic decay makes it possible to determine the shape of
the diffusive scattering.

Since the expression for $\langle s_{\bf 0} s_{\bf r} \rangle$ is only valid
for $r \gg 1$ we expect our evaluation to work for
$|{\bf p} - {\bf q}_j| \ll \pi$ (the extent of the first Brillouin zone).
With a reasonably fast ${\bf p}$-space decay of the peaks this restriction
will have negligible consequences. The conversion of the sum to
an integral will cause no trouble for a sufficiently large system, but
when making computer simulations on modestly sized lattices we must be
on guard for finite-size effects.

We have performed standard Monte Carlo simulations on $L \times L$
lattices for different sizes up to $L=900$. Periodic boundary conditions
are imposed and we demand $L$ to be divisible by 6 in order to avoid the
introduction of screw dislocations in the ground state \cite{Blote93}.
After initialising the lattice paramagnetically we equilibrate it at a
temperature $T$ (measured in units of $k_B / J$) by performing a suitable
number of MC steps per spin (MCSS).

Monitoring the excess energy per spin at $T=0$ as a function of MCSS (see
Fig.~\ref{Relax}) we find that the system after a short transient time
enters a regime of linear relaxation where it looses most of its excitational
energy. Qualitatively this regime corresponds to a `trivial' relaxation
during which the energy can be efficiently lowered by updating the
spin configuration locally. After a certain cross-over time which, as
demonstated on the inset of Fig.~\ref{Relax}, is proportional to $L$ a new
regime of roughly algebraic decay is entered as the annealing defects
generated during the quench are slowly eliminated and long-range
correlations are built up. Although the system at this point is
energetically very close to the ground state, we must take the length
of our runs to be considerably longer than this cross-over time in
order to obtain reliable data for the Bragg peaks in the structure factor.
To ensure that the MC data for different lattice sizes are of a
comparable quality we shall thus take the number of MCSS to be
proportional to $L$ and fix MCSS at 10,000 for $L=900$.

After thermalisation we calculate the structure factor $S({\bf p})$ for the
wave vectors in the
vicinity of ${\bf p} = {\bf q}_1$ which are compatible with translational
invariance, reinitialise, and make a new quench.
(Since only a small fraction of the simulation time
is spent in the regime of linear relaxation, as mentioned above, it
would not be advantageous to perform the simulations as sequential
heating runs. Instead, the method of repeated thermal quenches
employed here guarantees the independence of the different runs.)
For each value of $T$ and $L$ we average $S({\bf p})$ over 20 quenches.
Interference with the finite system size, however, manifests itself as a
kind of noise on the scale of the least allowable wave vector, even after
calculating the ensemble average. This ruggedness tends to obscure the shape
of the peak, wherefore we dispose of it by doing a mild coarse-graining
replacing each value of $S({\bf p})$ with the average of
itself and its four nearest neighbours. The result is a smooth peak as shown in
Fig.~\ref{Struc}.

The position of the Bragg peaks as calculated from their first moment compares
favourably with the theoretical $T=0$ result. For $T=0.5$ and $L=900$ we find
${\bf q}_1 = \pi \left( 1.33340(6), -0.00010(9) \right)
 \simeq \left( \frac{4\pi}{3},0 \right)$
with similar results for other temperatures and lattice sizes.

To avoid a circumstantial notation we focus our attention on one
of the two peaks in $S({\bf p})$, picking out the $j=1$ term of
Eq.~(\ref{TwinPeaks}) in the following. Shifting the
${\bf p}$-space origin to the centre of the peak we have
\begin{equation}
  S({\bf p}) \sim \int \mbox{\rm d}^2r e^{i{\bf p} \cdot {\bf r}} f(r)
             = 2\pi \int_0^{\infty} \mbox{\rm d} r \, r f(r) J_0(pr),
  \label{Bessel}
\end{equation}
where $J_0(x)$ is a spherical Bessel function of order zero.

Consider first the case of $T=0$ where Stephenson's result is represented
by $f(r) = 1/\sqrt{r}$. The integral can be written in terms of gamma
functions \cite{Gradshteyn} as
$S({\bf p}) \sim 2\pi \sqrt{2} p^{-3/2}
  \Gamma \left( \frac{3}{4} \right) /
  \Gamma \left( \frac{1}{4} \right)$ or
\begin{equation}
  S({\bf p}) \sim 3.0033 \, p^{-(2-\eta)}
  \label{Zero}
\end{equation}
with $\eta = \frac{1}{2}$.

This prediction is readily verified from our simulation data by computing a
circular average of $S({\bf p})$. 
The result for the $L=900$ lattice at $T=0$ is displayed in
Fig.~\ref{T0-circ}. As expected we find a power law behaviour
$S(p) \sim p^{-(2-\eta_{900})}$, valid in all but the immediate vicinity of
the peak. The exponent is $\eta_{900} = 0.45$.

Finite-size effects enter in two different ways. Firstly, the exponent deviates
slightly from the theoretical value $\eta = \frac{1}{2}$.
Secondly, the impossibility of reproducing exact divergences on a finite
lattice means that the power law fit breaks down near the peak centre $p=0$.
This can be interpreted as the effect of a large but finite correlation
length as described below.

At finite temperatures $T>0$ the system is no longer critical and is, as
opposed to the zero-temperature case, characterised by a finite correlation
length. The Stephenson expression for the decay of correlations in
this case \cite[Eq.~(1.31)]{Steph70a} can be rewritten as
\begin{equation}
  f(r) = \frac{e^{-r/ \xi}}{\sqrt{r}},
  \label{Sqrt}
\end{equation}
where the correlation length is given by $\xi^{-1} = - \ln \tanh \frac{1}{T}$ 
thus implying that $\nu = 1$ \cite{McCoy-Wu}.
Actually the oscillatory
factor is now proportional to $\cos \left( c(T) \frac{2 \pi r}{3}
\right)$, but since $c(T) \rightarrow 1$ as $T \rightarrow 0$ the
deviation from $\cos \left( \frac{2 \pi r}{3} \right)$ is of no
consequence for sufficiently low temperatures \cite{LowTemp}.
Note that Eq.~(\ref{Sqrt}) reduces to the $T=0$ result for
$\xi(T) \rightarrow \infty$ as it should.

Performing the $r$-integral \cite{Gradshteyn} we arrive at
\begin{equation}
  S({\bf p}) \sim 3.0033 \, p^{-(2-\eta)} g(p \xi),
  \label{Non-Zero}
\end{equation}
where the function $g(x)$, shown in Fig.~\ref{g(x)}, can be expressed as a
hypergeometric function \cite{Hypergeo}. Since $g(x) \rightarrow 1$ for
$x \rightarrow \infty$ the consistency with Eq.~(\ref{Zero}) is evident.

For high values of $\xi$ the expression for $S({\bf p})$ retains its pure
power-law form for all wave vectors $p$, since those allowed are bounded
from below
by $\frac{2\pi}{L}$. On the other hand we have $g(x) \sim x^{2-\eta}$ for
$x \rightarrow 0$, whence a lower value of $\xi$ makes $S({\bf p})$ tend
towards a constant for small wave vectors.

Our data for the circular averages of the diffusive scattering at
non-zero temperatures are shown in Fig.~\ref{Diffusive}, and the
agreement with the predictions of Eq.~(\ref{Non-Zero}) is seen to be
quite good. The curves shown are parametrised by the values of
$\xi(T)$ obtained from the exact expression given above, but as
shown in Table 1 these values are within 10 \% of the optimum
parameters found from a non-linear least-square fit. However, at very
low temperatures ($T \le 0.25$) when the deviation from a pure power
law decay becomes minute we are unable to reproduce the $T \rightarrow
0$ divergence in $\xi(T)$ due to the finite size of our lattices.

We now turn our attention to a scaling analysis of the data
obtained for various lattices. According to the standard theory of finite-size
scaling \cite{Barber} the spin-spin correlation function should scale
with system size $L$ as
$\langle s_{\bf 0} s_{\bf r} \rangle = L^{-2\beta / \nu} \,
 \mbox{\rm f} \left( \frac{r}{L},t L^{1/\nu} \right)$, where $t=\frac{T-T_c}{T_c}$ is
the reduced temperature. When applying this scaling ansatz to the TAI
model two caveats must be taken into account.
Firstly, $t$ is not a suitable measure of the dimensionless temperature
in a model with $T_c = 0$. Secondly, the critical exponent
$\beta$, which is defined in terms of the spontaneous magnetisation below the
critical temperature, is a very doubtful parameter indeed, since its domain
of definition simply does not exist.

We shall dispose of the first complication
by appealing to the Coulomb gas description given in
Ref.~\cite{Blote84} according to which a temperature induced
excitation from the TAI model ground state is equivalent to the
formation of a vortex in the associated SOS-model. These vortices
correspond to all three spins on an elementary triangle being aligned
and accordingly have the Boltzmann weight $\mbox{\rm e}^{-2/T}$
when expressed by our dimensionless variables. The second problem is
easily eliminated, since an inspection of Kadanoff's block spin
argument \cite{Kadanoff} reveals that the factor $L^{-2 \beta / \nu}$
is only conventional and may be replaced by $L^{- \eta}$ without ever
referring to $\beta$.

Finally, we define a sublattice independent correlation function
$\Gamma(r,T) = \langle s_{\bf 0} s_{\bf r} \rangle (3 \delta_{i_r,A} - 2)$,
where $i_r$ is the label of the sublattice (we fix the origo so that
$i_0 = A$), i.e., by
eliminating the dependence on the weight factor. The finite-size
scaling hypothesis now assumes the form
\begin{equation}
  \Gamma(r,T) = L^{-\eta} \, \mbox{\rm f} \left(
                \frac{r}{L},\mbox{\rm e}^{-2/T} L^{1/\nu}
                \right).
\end{equation}
Note that $\eta$ is the correlation function exponent pertinent to the
staggered field. Albeit the scenario of Ref.~\cite{Blote84} opens the
possibility of three diffent spin waves, each with its own value of
$\eta$, the period-2 wave would only be present if we were to include an
external magnetic field, and the period-3 wave would only be relevant
if the coupling constants were sublattice dependent.

At zero temperature we have, from Eq.~(\ref{Steph}), that
$\Gamma(r,0) \sim r^{-\eta}$
for $r \gg 1$, whence
\begin{equation}
  \mbox{\rm f} (x,0) \sim x^{-\eta}
\end{equation}
for values of $x = \frac{r}{L}$ satisfying $0 \ll x \ll 1$. As witnessed
by Fig.~\ref{finsize-00} this asymptotic behaviour of $\mbox{\rm f}(x,0)$
is in fact nicely brought out for {\em all} values of $x$ shown on the
graph, at least for the
small lattices. When increasing the lattice size we first see an unexpected
fall-off for large $x$ and eventually, for $L=900$, even a clear tendency
for $\Gamma$ to ramify in three individual curves corresponding to the
three sublattices.

The latter observation could be taken to imply that the largest lattices
have been insufficiently thermalised to correctly bring out the $T=0$
correlations for distances above some one hundred lattice constants. To
check this hypothesis we have produced some extra runs for the $L=300$
lattice, using a different number of MCSS than in our main series of data.
In Fig.~\ref{breakdown} we have redisplayed the $L=300$ data as above
along with the results obtained when using 10\%, 30\% and 300\% of the nominal
number of MCSS. The results clearly corroborate our suspicion, that both
the anomalous fall-off and the sublattice upsplitting are indeed effects
of an insufficient thermalisation time. Considerations on currently available
CPU time, however, prevented us from performing longer runs than the ones
reported here.

The dependence of the scaling function $\mbox{\rm f} (x,y)$ on its
second variable can be examined by plotting correlation data for a
range of non-zero temperatures
and different lattice sizes in a series of curves with a fixed value
$x = x_0$ \cite{Landau75}.
In Fig.~\ref{finsize-T} we show the plot of $\mbox{\rm f} (x_0,y)$
versus $y = \mbox{\rm e}^{-2/T} L^{1/\nu}$, and it is
found that the choice $\nu = 1.0 \pm 0.1$ makes the data collapse on
distinct curves parametrised by the different values of $x_0$.
As expected from the above discussion on insufficient thermalisation
the quality of the $x_0 = \frac{1}{6}$ curve is
inferior to that of the other branches of the graph.

Note, that for sufficiently low $y$, i.e., at low temperatures, $\mbox{\rm f} (x_0,y)$ is
constant for a fixed value of $\frac{r}{L}$. Thus
$\Gamma(r,T) \sim L^{-\eta} \sim r^{-\eta}$, whence the zero temperature
algebraic decay $\Gamma(r,0) \sim r^{-\eta}$ is in fact valid even for
small finite values of $\mbox{\rm e}^{-2/T} L^{1/\nu}$.


In summary, we have shown that Stephenson's expression
(Eq.~\ref{Steph}) can indeed be applied to all lattice directions in
the way suggested by Ref.~\cite{Blote84}. Moreover, we have confirmed
the validity of the appropriate generalisation to small non-zero
temperatures by exhibiting the agreement with our Monte Carlo data.
Finally, a finite-size scaling analysis demonstrated that the correct
way of approaching the $T=0$ critical point is through the temperature
parameter $t = \mbox{\rm e}^{-2/T}$.


The authors wish to thank Ole Mouritsen for having originally brought the
TAI model to our attention.



\begin{figure}
  \caption{The $L \times L$ triangular lattice spanned by
           ${\bf a}_{1,2} = (\pm \frac{1}{2},\frac{\sqrt{3}}{2})$
           can be divided into three interpenetrating sublattices labelled
           $A$, $B$ and $C$. Sublattice $A$ is spanned by
           ${\bf a}_{1,2}^A = (\pm \frac{3}{2},\frac{\sqrt{3}}{2})$.
           When imposing periodic boundary conditions, $L$ must be divisible
           by six in order to preserve this sublattice structure and avoid
           the introduction of screw dislocations in the ground state.
           Note, that next-nearest neighbours belong to the same sublattice,
           whence a ferromagnetic coupling between them tends to align the
           spins on each sublattice and thus lift the ground state degeneracy.}
  \label{Lattice}
\end{figure}

\begin{figure}
  \caption{MC simulation of the TAI model on a $90 \times 90$ lattice clearly
           shows the difference between the appearance of the paramagnetic
           initial state (left) and the labyrinthine patterns of the
           locally ordered
           ground state (right), which is reached after 1000 MC steps per spin.
           Here, spin up (down) is represented by the presence (absence) of a
           dot.}
  \label{Pattern}
\end{figure}

\begin{figure}
  \caption{A point in the reciprocal lattice spanned by
           ${\bf b}_{1,2} = 2\pi (\pm 1,\frac{1}{\sqrt{3}})$, with its six
           nearest neighbours. The inner hexagon is the first Brillouin zone.
           Variational mean field theory predicts that the Bragg points
           describing the continuous transition to an ordered phase be located
           according to the Lifshitz criterion, i.e., that they be points of
           high symmetry in the first Brillouin zone. Apart from the origo,
           which characterises a genuine long-range order, the possibilities
           satisfying this symmetry demand are marked by a dot in the figure.
           Imposing the additional condition that the Landau free energy be
           minimal, we are left with the two Bragg peaks at ${\bf q}_1$ and
           ${\bf q}_2$. Fourier transformation of the pair-correlation
           function confirms the mean field scenario.}
  \label{Brillouin}
\end{figure}

\begin{figure}
  \caption{The main graph shows the excess energy per spin at $T=0$, measured
           relative to the ground state, as a function of Monte Carlo time for
           a range of different lattice sizes $L=300$, 600, 900, 1200 and 1500.
           After a short transient time the system enters a regime of linear
           relaxation terminating in a cross-over to an algebraic decay as the
           ground state is approached. As demonstrated in the inset this
           cross-over time is proportional to $L$.}
  \label{Relax}
\end{figure}

\begin{figure}
  \caption{Close-up of the structure factor in the vicinity of the Bragg
           point ${\bf q}_1 = \left( \frac{4\pi}{3},0 \right)$ for a system
           of size $L=900$ at temperature $T=0.5$. An average over 20
           independent quenches as well as a mild coarse-graining have been
           performed.}
  \label{Struc}
\end{figure}

\begin{figure}
  \caption{Circular average showing the line shape of the $T=0$ Bragg peak
           simulated on a $L=900$ lattice. The exponent of the algebraic
           decay is $-(2 - \eta_{900}) = -1.55$.}
  \label{T0-circ}
\end{figure}

\begin{figure}
  \caption{The function $g(x)$ \protect \cite{Hypergeo} controlling the
           deviation of the
           diffusive scattering from a pure power law form. From the
           asymptotic behaviour we infer the constancy of $S({\bf p})$ at
           small values of $x = p \xi$ (using $g(x) \sim x^{2-\eta}$ for
           $x \rightarrow 0$) and the unchanged form at large ones (using
           $g(x) \rightarrow 1$ for $x \rightarrow \infty$).}
  \label{g(x)}
\end{figure}

\begin{figure}
  \caption{Fits to the shape of the diffusive scattering at temperatures
           $T=0.25$, 0.35, 0.40, 0.45, 0.50 and 0.60 as indicated by
           the labels.
           For clarity the graphs have been shifted along the $S(p)$-axis.}
  \label{Diffusive}
\end{figure}

\begin{figure}
  \caption{Finite-size scaling at $T=0$. The graphs for $\mbox{\rm f}(x,0)$
           versus $x$ for different system sizes
           ($L=90$, 150, 300, 600 and 900 as labelled) collapse on a universal
           curve. The breakdown of scaling at large $x$ is due to an
           insufficient thermalisation time.}
  \label{finsize-00}
\end{figure}

\begin{figure}
  \caption{Replotting the $L=300$ data for different thermalisation times
           (the label indicates the percentage of the nominal number of MCSS)
           we infer that the deviation from the universal scaling function
           $\mbox{\rm f}(x,0) \sim x^{-\eta}$ could be ameliorated by making
           longer runs.}
  \label{breakdown}
\end{figure}

\begin{figure}
  \caption{Finite-size scaling at $T>0$. The graphs for $\mbox{\rm f}(x_0,y)$
           versus $y$ fall on distinct branches according to the value of the
           parameter $x_0 = r/L$. For clarity the different branches have
           been shifted along the ordinate.}
  \label{finsize-T}
\end{figure}


\begin{table}
 \begin{center}
  \begin{tabular}{|c||c|c|c|c|c|c|}     \hline
    $T$             & 0.25 & 0.35 & 0.40 & 0.45 & 0.50 & 0.60   \\ \hline
    $\xi$ (fit)     &  172 & 124 &  82 &  48 &  29 &  15   \\ \hline
    $\xi$ (exact)   & 1490 & 152 &  74 &  43 &  27 &  14   \\ \hline
  \end{tabular}
  \caption{Correlation lengths $\xi$ (in units of the lattice constant) for a
           $L=900$ lattice as a function of temperature $T$.}
  \label{CorrLength}
 \end{center}
\end{table}


\begin{references}
\bibitem {LowTemp}
	For $T = 0.6$ which is the highest temperature used in our
	simulations $c(T) = 0.99930$ implying that sublattices begin
	to mix up for $r \ge 1425$. Since our largest lattices are of
	size $L = 900$ the deviation of $c(T)$ from 1 is indeed negligible.
\bibitem {Hypergeo}
        From \protect \cite{Gradshteyn} we have
        $g(x) = \frac{1}{4 \sqrt{\pi}} \Gamma^2 \left( \frac{1}{4} \right)
                (1+x^{-2})^{-3/4}
                F\left( \frac{3}{4},-\frac{1}{4};1;\frac{1}{1+x^{-2}} \right),$
        where the Gaussian hypergeometric function
        $F(\alpha,\beta;\gamma;z)$ is defined by the series
        $\sum_{n=0}^{\infty}
        \frac{(\alpha)_n (\beta)_n}{(\gamma)_n} \frac{z^n}{n!}$,
        which is convergent for $|z| \le 1$.
        Since the convergence of $F$ is extremely slow for the values
        $z \simeq 1$ useful at low temperatures, it is advantageous to employ a
        translation formula for the hypergeometric function \cite{Gradshteyn}
        yielding
        $g(x) = (1+x^{-2})^{-3/4} \left( F \left(
                \frac{3}{4},-\frac{1}{4};\frac{1}{2};\frac{1}{1+x^2} \right) +
                \frac{k}{\sqrt{1+x^2}} F \left(
                \frac{1}{4},\frac{5}{4};\frac{3}{2};\frac{1}{1+x^2} \right)
                \right),$
        where
        $k = 2 \Gamma^2 \left( \frac{1}{4} \right) /
               \Gamma^2 \left( -\frac{1}{4} \right) \simeq 1.094,$
        whereafter the two new hypergeometric functions can be expanded to
        low order.
\end{references}
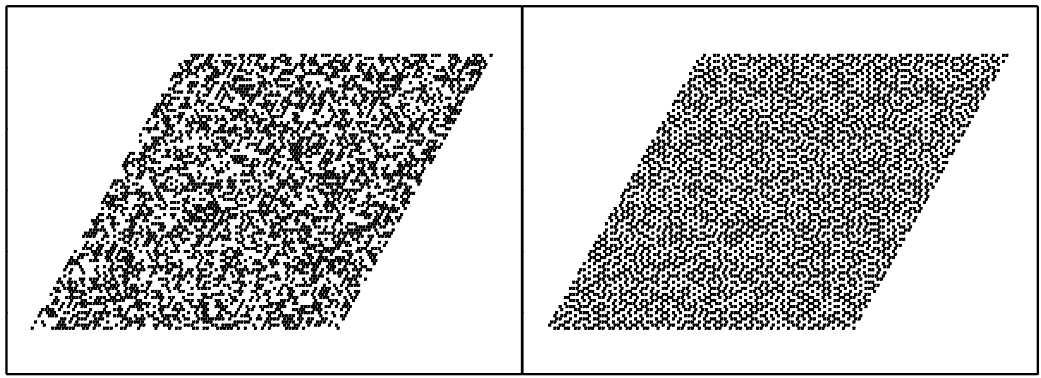
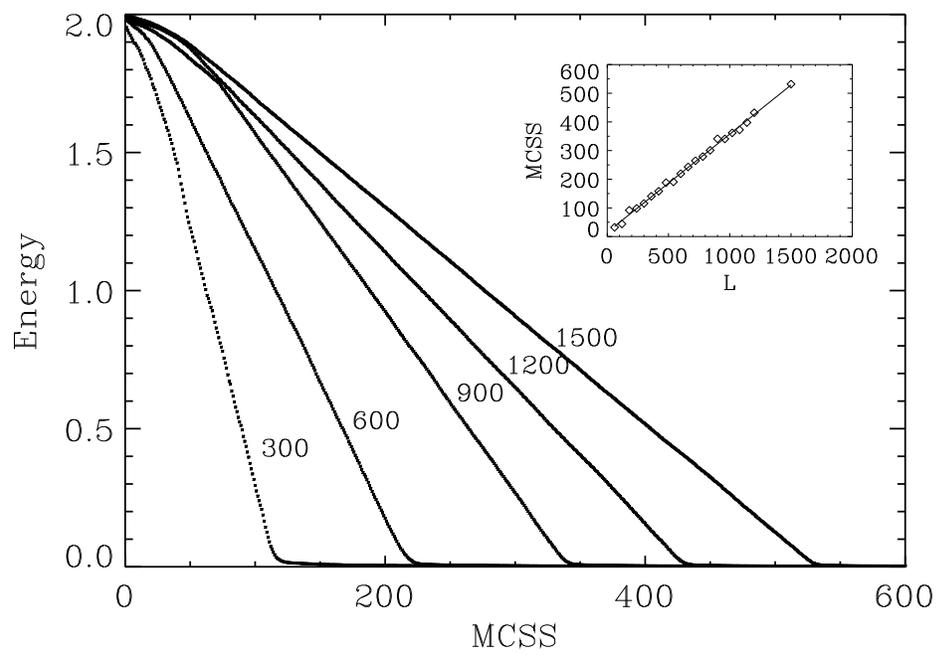
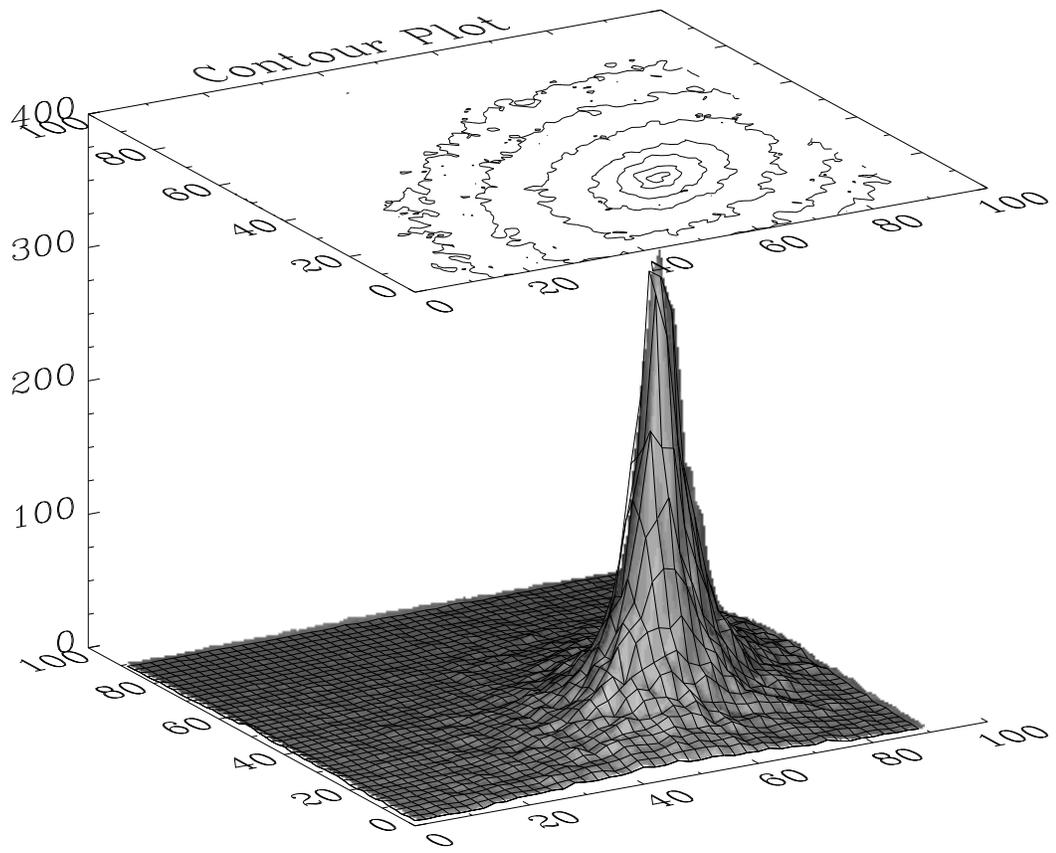
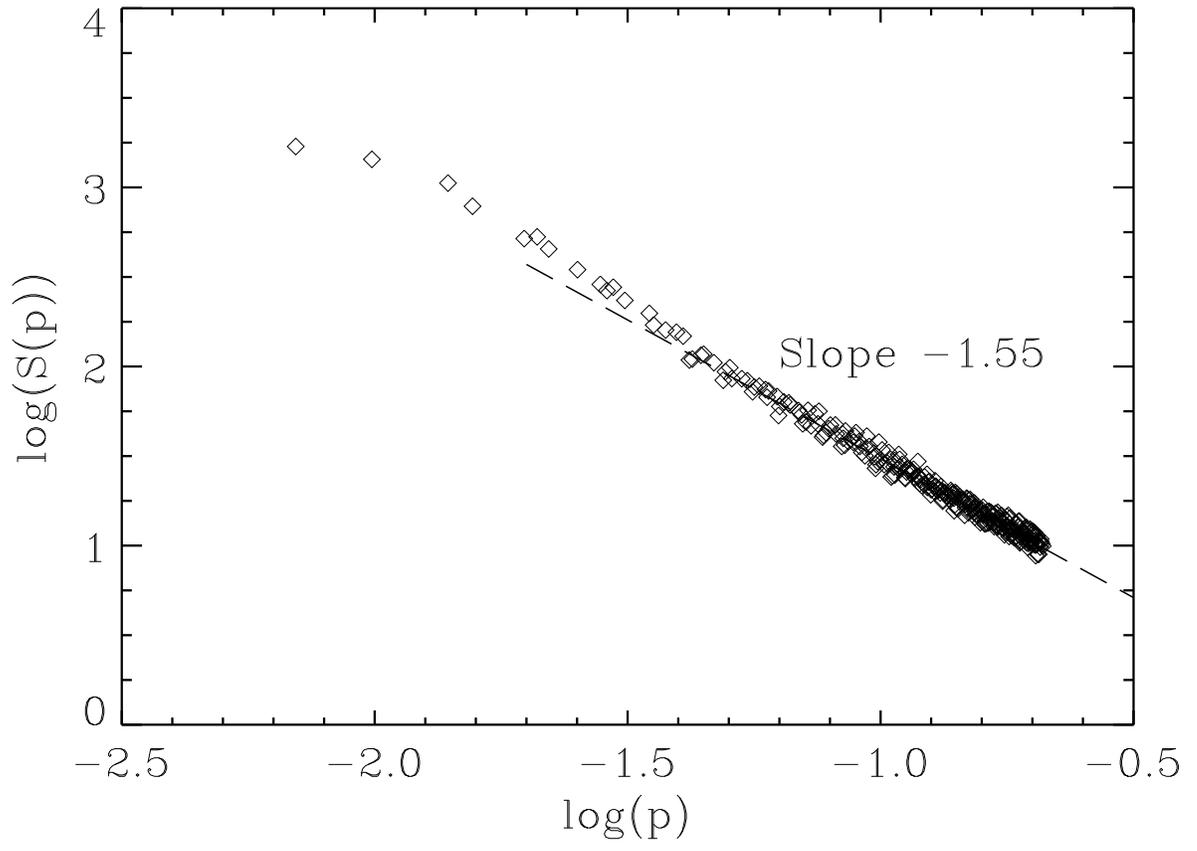
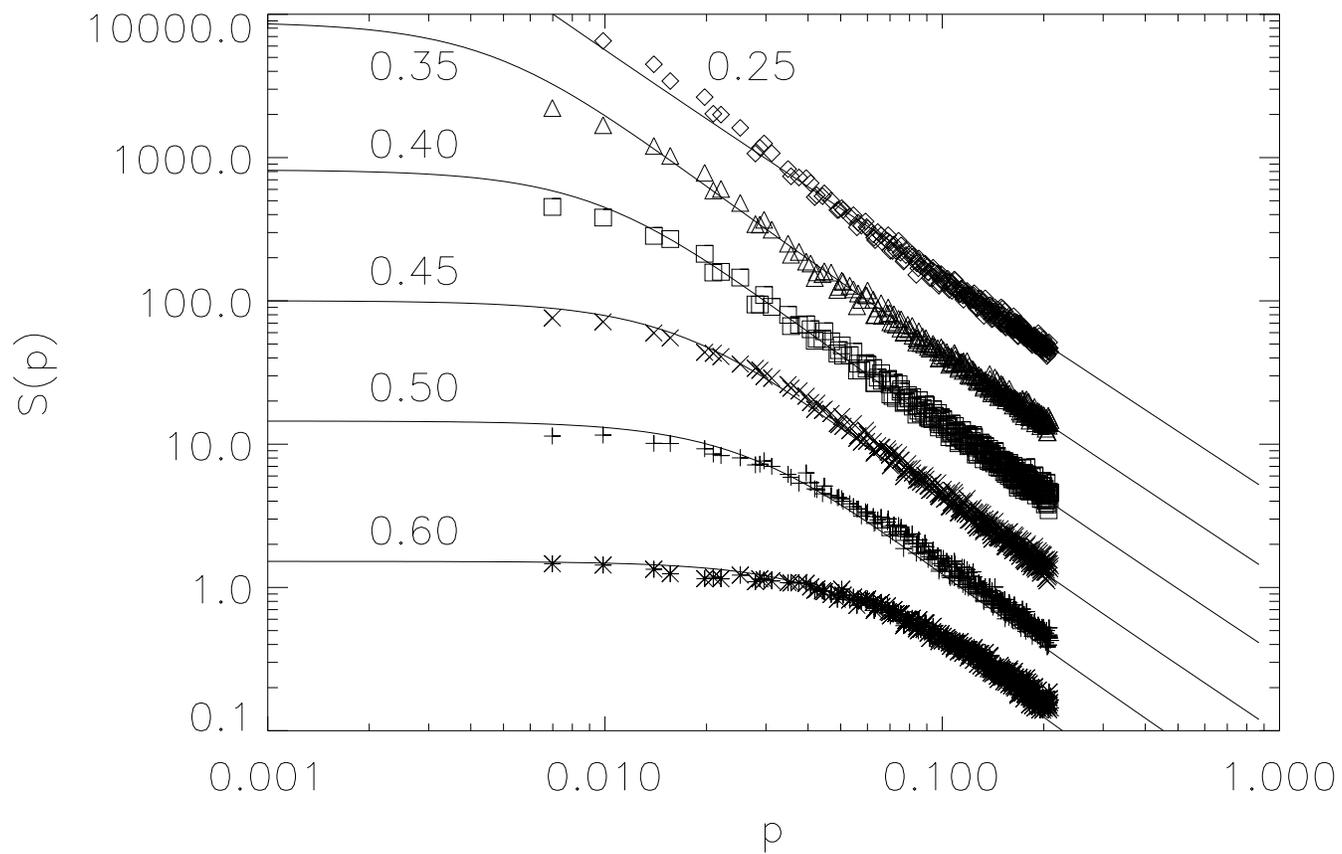
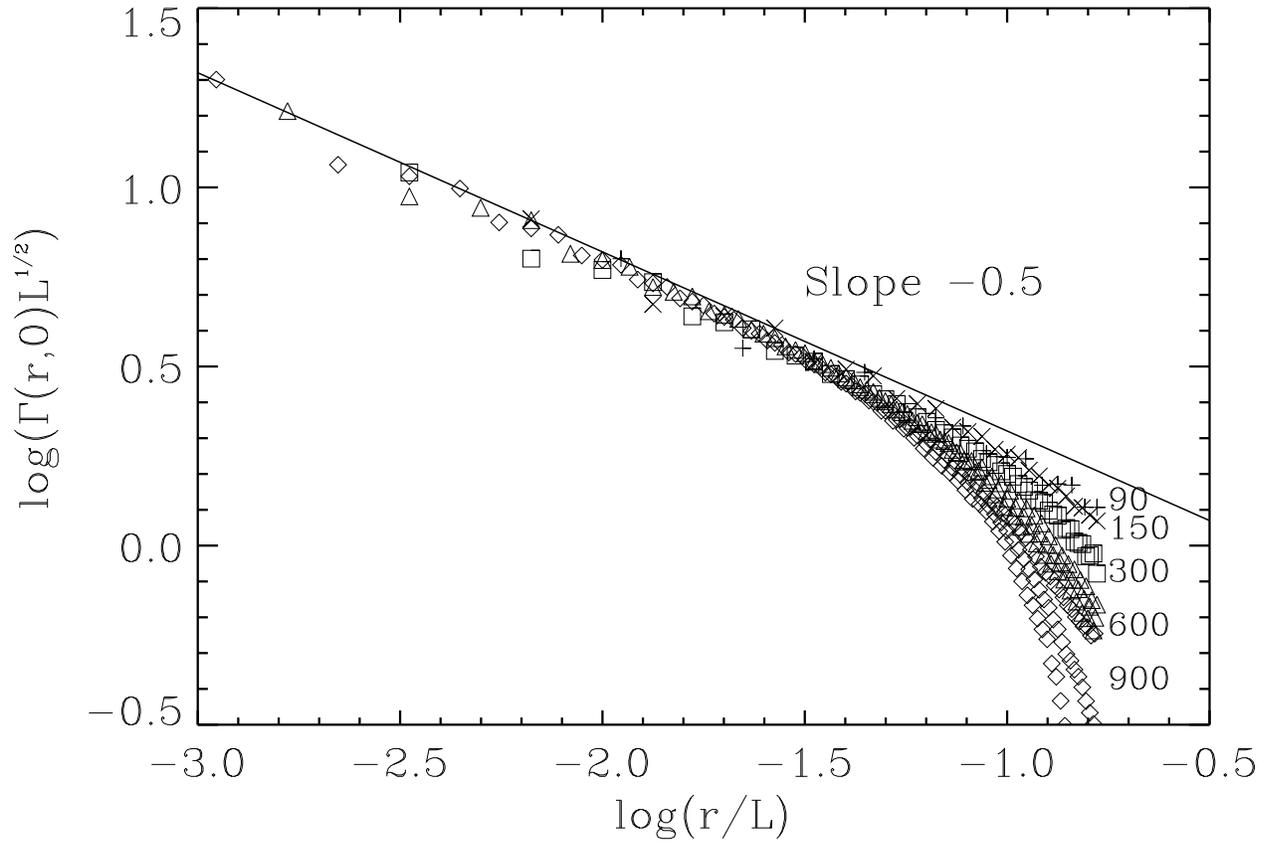
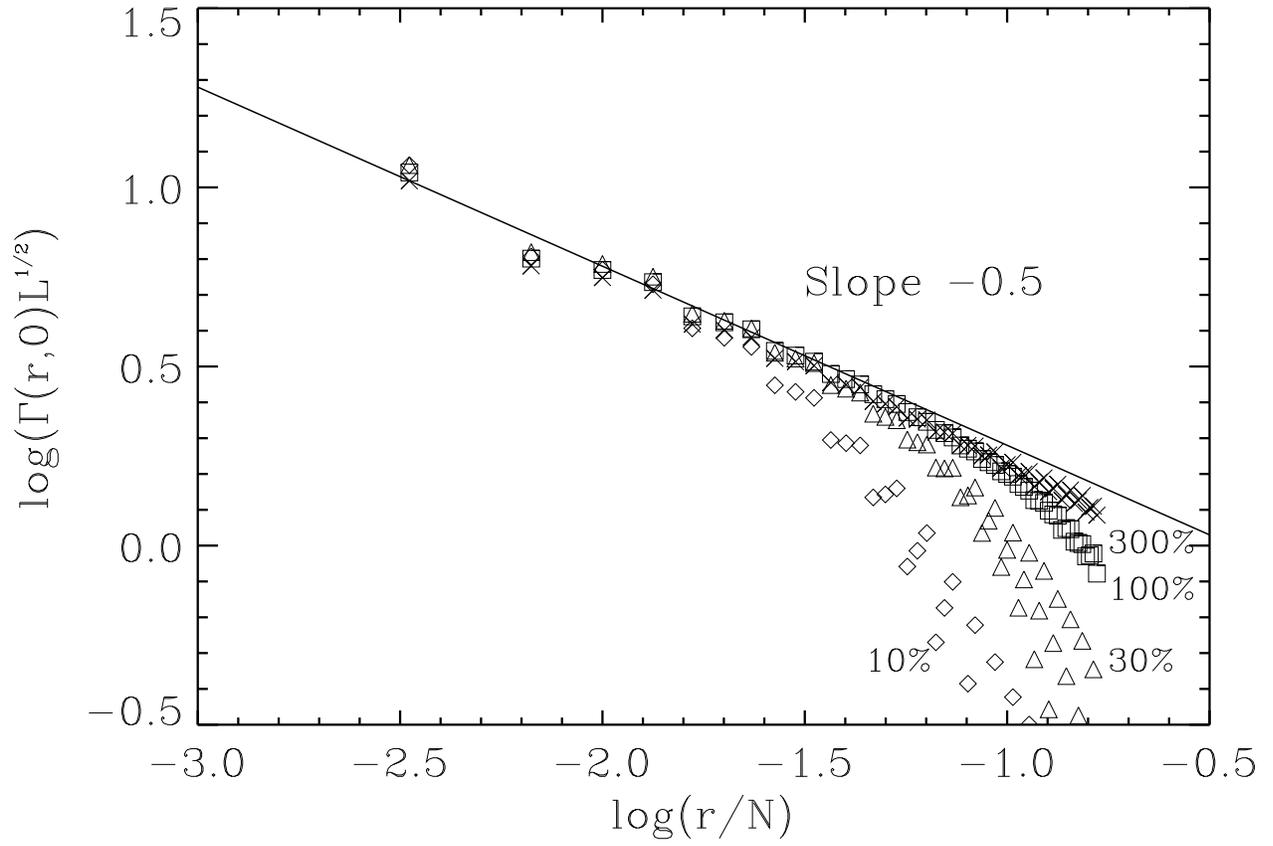
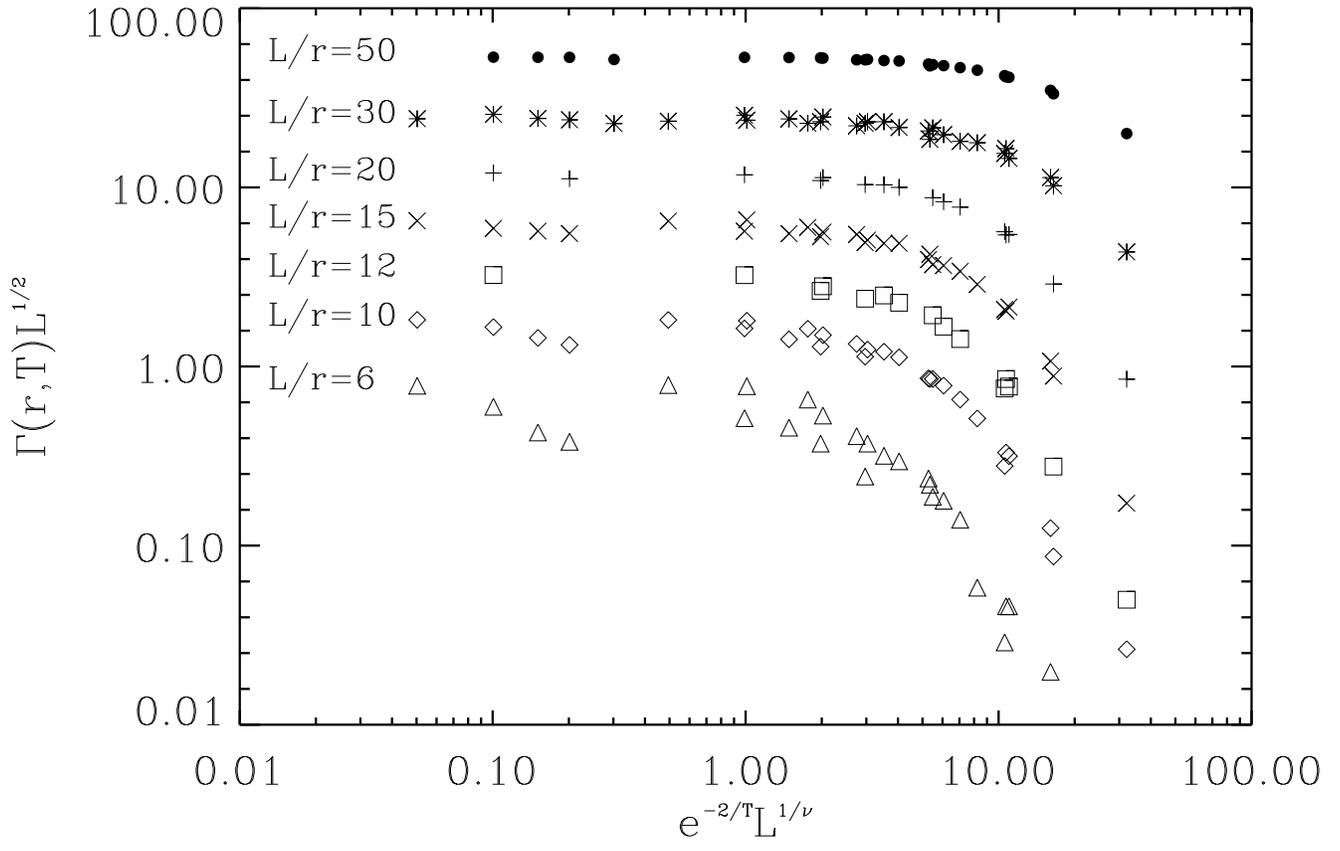
\end{document}